# Neutron diffraction on antiferromagnetic ordering in single-crystal BaFe$_2$As$_2$

Y. Su[1*], P. Link[2], A. Schneidewind[3], Th. Wolf[4], P. Adelmann[4], Y. Xiao[5], M. Meven[2], R. Mittal[1], M. Rotter[6], D. Johrendt[6], Th. Brueckel[1,5], M. Loewenhaupt[3]

[1]*Juelich Centre for Neutron Science, IFF, Forschunszentrum Juelich, Outstation at FRM II, Lichtenbergstr. 1, D-85747 Garching, Germany*
[2]*FRM II, Technische Universitaet Muenchen, Lichtenbergstr. 1, D-85747 Garching, Germany*
[3]*Institut fuer Festkoerperphysik, Technische Universitaet Dresden, D-01062 Dresden, Germany*
[4]*Forschungszentrum Karlsruhe, Institut fuer Festkoerperphysik, D-76021 Karlsruhe, Germany*
[5]*Institut fuer Festkoerperforschung, Forschungszentrum Juelich, D-52425 Juelich, Germany*
[6]*Department Chemie und Biochemie, Ludwig-Maximilians-Universitaet Muenchen, Butenandtstrasse 5-13 (Haus D), D-81377 Muenchen, Germany*

Abstract

Neutron diffraction experiments have been carried out on a Sn-flux grown BaFe$_2$As$_2$ single crystal, the parent compound of the A-122 family of FeAs-based high-$T_c$ superconductors. A tetragonal to orthorhombic structural phase transition and a three dimensional long–range antiferromagnetic ordering of the iron magnetic moment, with a unique magnetic propagation wavevector **k** = (1, 0, 1), have been found to take place at ~90 K. The magnetic moments of iron are aligned along the long **a** axis in the low temperature orthorhombic phase (*Fmmm* with **b**<**a**<**c**). Our results thus demonstrate that the magnetic structure of BaFe$_2$As$_2$ single crystal is the same as those in other A-122 iron pnictides compounds. We argue that the tin incorporation in the lattice is responsible for a smaller orthorhombic splitting and lower Néel temperature $T_N$ observed in the experiment.

PACS numbers: 75.25.Ha; 75.25.+z, 75.40.Cx, 75.50.Ee

Recent discovery of iron arsenic based superconductors LaFeAsO$_{1-x}$F$_x$ [1-2] (La-1111) has generated tremendous interests in searching for new high transition temperature ($T_C$) superconductors and the underlying mechanisms. Superconductivity with $T_C$ values up to 56 K has been subsequently discovered in other RE-1111 (RE=Sm, Ce, Nd, Pr) compounds [3]. The second family of FeAs-based superconductors, A$_{1-x}$K$_x$Fe$_2$As$_2$ ($x$ ~ 0.4) (A=Ba, Ca, Sr) (A-122) with $T_C$=38 K has been discovered more recently [4]. Two common aspects of both RE-1111 and A-122 families have been experimentally revealed. The first one is that their parent compounds likely display an antiferromagnetic (AF) ordering of iron spins at low temperatures and are accompanied by lattice distortions. The AF ordering and the corresponding structural phase transitions have been strongly supported by neutron powder diffraction experiments on LaFeAsO [5,6], NdFeAsO [7,8], CeFeAsO [9] and BaFe$_2$As$_2$ [10,11]. Another aspect is that superconductivity could emerge upon electron- or hole-doping via chemical substitutions of the parent compounds. These two features are in strong reminiscence to the behaviors in the cuprates.

Despite of the apparent similarities between the newly discovered iron pnictides and cuprates, the fundamental issues concerning the nature of magnetic ground states in the parent compounds and possible role of spin fluctuations in the pairing mechanism of superconductivity are far from clear. From nominal chemical composition (e.g. La$^{3+}$Fe$^{2+}$As$^{3-}$

$O^{2-}$) and the local configuration of FeAs$_4$ tetrahedron in iron pnictides, a high-spin Fe$^{2+}$ seems a likely assignment. However, the magnetic moments of iron reported in recent neutron powder diffraction investigations are very small, for instance, 0.35 $\mu_B$ in LaFeAsO [5], 0.25 $\mu_B$ in NdFeAsO [8] and 0.87 $\mu_B$ in BaFe$_2$As$_2$ [11]. This raises questions on the exact nature of the electronic and spin states of iron in these compounds. Different scenarios have thus been suggested to explain the smallness of the iron magnetic moments [12]. For instance, this may be related to possible presence of magnetic frustrations due to strong competitions between the nearest and the second-nearest neighbor exchanges interactions, or this is due to a prominent itinerant character of iron spins, and so on. In cuprates, both AF ordering and incommensurate spin fluctuations are quasi two-dimensional due to a large inter-layer distance between neighboring Cu-O planes. The low dimensionality in cuprates weakens the magnetism and paves the way for high-$T_C$ superconductivity. Iron pnictides compounds bear a similar layered structure to cuprates, therefore, it is naturally expected that the low dimensionality might also play an important role here. The recent success in single crystal growth of A-122 compounds [13-15] has provided unprecedented opportunities to clarify these important issues, in particular, via single-crystal neutron scattering. In this article we report comprehensive magnetic neutron diffraction experiments on a Sn-flux grown BaFe$_2$As$_2$ single crystal. A long-range antiferromagnetic ordering has been observed to occur below $T_N$ ~ 90 K with a unique magnetic propagation wavevector **k** = (1, 0, 1) (the lattice parameters determined at 3 K: a = 5.587 Å, b = 5.558 Å, c = 13.022 Å). The iron magnetic moments are aligned along the orthorhombic **a** axis.

A single crystal of BaFe$_2$As$_2$, with the dimension of about 4×4×0.2 mm was grown from the Sn flux as reported elsewhere [14]. It is thin platelet-like and with a shiny (0, 0, 1) surface. The thorough characterizations via specific heat, resistivity, magnetic susceptibility and chemical composition analysis have been carried out on the crystals obtained from the same batch. Similar physical properties to those in the crystals reported in Ref [14] have been obtained. The EDX composition analysis of several specimens of this batch resulted an average composition of Ba$_{0.95}$Sn$_{0.05}$Fe$_2$As$_2$ and we expect ~5% tin to be incorporated in the crystal structure. An anomaly due to magnetic ordering and structural phase transition is clearly visible at ~90 K in heat capacity. The lower magnetic transition temperature (comparing to $T_N$ ~140 K in polycrystalline samples [10]) has also been observed in BaFe$_2$As$_2$ single crystals grown in tin-flux elsewhere [14,16]. Single crystals obtained from the same batch were thoroughly grounded into fine powders. High-resolution X-ray powder diffraction experiment has been carried out using a Huber G670 Guinier imaging plate diffractometer with Cu-K$_{\alpha 1}$ radiation and a Ge-111 monochromator. The sample is cooled to low temperatures with a closed-cycle displex. The crystal structure has been determined to be tetragonal (*I4/mmm*) at high temperatures and orthorhombic (*Fmmm*) below ~ 90 K via the Rietveld refinement. The lattice parameters have been precisely determined as well, as shown in Fig. 1. The tetragonal to orthorhombic phase transition can be clearly indicated from an abrupt change of the orthorhombic distortion parameter $P$ (= (a-b)/(a+b)) at ~ 90 K (Fig. 1(b)) Note that $P$ is only about 0.25% at 3 K, which is substantially reduced from ~0.4% obtained from the polycrystalline samples [10].

The single-crystal neutron scattering experiment has been carried out at PANDA, a high-flux cold-neutron triple-axis spectrometer at FRM II. $k_i$ and $k_f$ from the PG002 monochromator and analyzer have been set to 2.662 Å$^{-1}$. A 6 cm pyrolytic graphite (PG) filter has been used to reduce high-order contaminations. The high resolution measurement has been undertaken with the collimation configuration 40'-15'-15'-open, while the integrated intensities were measured without any collimations in the beam path. The experiment has also been carried out at hot-neutron four-circle single-crystal diffractometer HEIDI at FRM II, with the neutron wavelength chosen at 1.16 Å. The crystal has been placed on an Al holder with strain-free mounting via UHV grease and cooled via a closed-cycle displex at both PANDA and HEIDI. BaFe$_2$As$_2$ undergoes a tetragonal-orthorhombic structural phase transition (*I4/mmm→Fmmm*) at low temperatures. Therefore, an orthorhombic notation with the shortest axis defined as **b** (**b**<**a**<**c**) will be adopted throughout this paper.

Both instruments were switched to the high Q-resolution mode with fine collimations in order to resolve the small orthorhombic splitting at low temperatures. However, the *d*-space splitting due to the formation of the orthorhombic twins is not resolvable from any longitudinal scans taken at HEIDI and PANDA. This is due to a combined effect from the limited instrument resolution and the smallness of the orthorhombic splitting. Nonetheless, as shown in Fig. 2(a), a triple splitting of the rocking curve of (2, 2, 0) can be clearly observed at 3 K, when the (*h*, *k*, 0) reciprocal plane is aligned in the horizontal scattering plane. This is the strongest indication of the occurrence of the expected tetragonal to orthorhombic phase transition. The temperature dependence of the rocking curve width of (4, 0, 0) and (2, 2, 0) upon both cooling and heating is shown in Fig. 2(b), indicating that the structural phase transition indeed takes place at ~ 90 K, while the width of (0, 0, 8) shows no changes in the corresponding temperature range. The longitudinal scan along the *h*-direction at (-4, 0, 0) measured with the high Q-resolution mode at PANDA is slightly asymmetric with an apparent shoulder $h \approx -3.980$, as shown in Fig. 2(c). The best fitting of the (-4, 0, 0) peak has been achieved with two Gaussians centered at $h = -4.003$ and $-3.980$, respectively. Therefore, the population ratio of the corresponding domains connected to the (0, *k*, 0) and (*h*, 0, 0) twins can be estimated to be 5.6:1.

Comprehensive search for all possible magnetic reflections has been carried out in the (*h*, 0, *l*) reciprocal plane (Fig. 3(a)) at PANDA. Magnetic Bragg peak intensities have been observed at (*h*, 0, *l*)$_M$ with *h*, *l* = odd (subscript M here refers to magnetic and N to nuclear), consistent with the earlier observation on powders [11]. As shown in Fig. 3(b), the AF reflection (-1, 0, -3)$_M$ has clearly been observed at 3 K. Note that (-1, 0, -3)$_N$ is forbidden in *Fmmm*. The temperature dependence of the magnetic intensity of (-1, 0, -3)$_M$ (see Fig. 3(c)), measured both upon heating and cooling, indicates that the Sn-flux grown BaFe$_2$As$_2$ single crystals undergo a magnetic phase transition with $T_N$ at 90 K. At T = 3 K scans have been performed around (-1, 0, -3)$_M$ along the *h*-direction in reciprocal space (Fig. 4(a)) with and without the PG filter. Without the PG filter, the intensities in the scan is dominated by the high-order contributions from the nuclear reflection (0, -2, -6)$_N$ from the most populated (0, *k*, 0) twin. The corresponding scans along the *l*-direction have been also undertaken to ensure that all reflections from the twinned domains reside at integer *l* positions. The position of (-1, 0, -3)$_M$ is apparently shifted away from the maximum of (0, -2, -6)$_N$ and is roughly at (-0.995, 0, -3). This strongly indicates that the magnetic ordering is only related to the (*h*, 0, 0) twin. Therefore, our data obtained on this crystal strongly supports a unique magnetic propagation wave vector **k** = (1, 0, 1).

The magnetic neutron scattering is only sensitive to the magnetization perpendicular to **Q**, therefore, it can be simplified as, $I_M(Q) \propto |f_M(Q)|^2 \cdot \sin^2\alpha$, $f_M(Q)$ is the magnetic form factor of iron, $\alpha$ is defined as the angle between **Q** (the scattering vector) and the magnetic moment direction, other trivial scale factors are ignored here. For magnetic reflections with the same |**Q**|, the intensities are thus only proportional to $\sin^2\alpha$. Convincing evidence for the moments to be aligned along the **a** direction can be drawn from the comparison of the intensity of the $(1, 0, 7)_M$ and $(3, 0, 1)_M$ magnetic reflections (Fig. 4(b)). The factor $\sin^2\alpha$ is 0.02 and 0.90 for $(3, 0, 1)_M$ and $(1, 0, 7)_M$ respectively, note that both reflections have almost equal $|Q|$ and therefore identical form factors. To refine the magnetic structure, a set of nuclear and magnetic reflections in the ($h$, 0, $l$) reciprocal plane have been measured at 3 K. The integrated intensities were obtained from the θ-2θ scans in order to properly correct the Lorentz factor for a triple-axis spectrometer. A least-squares refinement has been performed with the FULLPROF package. A common scale factor has been used in the refinement of both nuclear and magnetic structure. The best fit from the total 12 magnetic reflections has been obtained with $R$-factor at 9.8%, based on a model with a unique modulation wavevector **k** = (1, 0, 1) and with the moment along the **a**-axis. The saturated moment has been determined to be 0.99(6) $\mu_B$ assuming a domain population ration of 5.6:1. The magnetic structure of $BaFe_2As_2$ can thus be described as shown in Fig. 4(c). The magnetic moments of iron are arranged anti-parallel to the neighboring ones along orthorhombic ***a***- and ***c***-axis, while they are parallel along ***b***-axis. The peak widths of magnetic reflections are resolution-limited along both ***a*** * and ***c*** * directions resulting in an estimated correlation length of at least 300 Å. This indicates that the magnetic structure in $BaFe_2As_2$ is three-dimensional and long-range ordered, despite of the large inter-layer distance between the neighboring Fe-As planes. This is similar to the case in two-dimensional $Na_{0.75}CoO_2$, the parent compound of Na-Co based superconductors, where three-dimensional spin fluctuations have actually been revealed by inelastic neutron scattering [17]. It worth to note that the magnetic structure of single-crystal $BaFe_2As_2$ reported here is thus same as those in the polycrystalline counterpart, as well as in $SrFe_2As_2$ [19] and $CaFe_2As_2$ [20] compounds.

At high temperatures, the parent compounds of both RE-1111 and A-122 families show the behavior of Pauli paramagnetism, as demonstrated from magnetic susceptibility measurements. The lattice distortions introduced from the structural phase transitions may eventually alter the electronic and spin states of Fe via subtle changes of the surrounding environment, for example, changes of strength and/or symmetry of crystal field. These subtle changes may pave the way for the emergence of magnetism. In addition, the distances from Fe to two nearest neighbors in the *a-b* plane become slightly different in the orthorhombic phase, which may help to release possible magnetic frustrations and favor anisotropic nearest neighbor exchange interactions and thus strongly influence magnetic ordering. Therefore, it has been generally believed that the magnetic ordering in iron pnictides is strongly coupled to the structural transition to a lower symmetry phase. Our results have indeed demonstrated that both magnetic and structural phase transitions take place almost at the same temperature. We argue that much smaller orthorhombic splitting and lower $T_N$ than those observed in parent polycrystalline samples are due to the Sn incorporation in the lattice. Even with only a small amount of Sn in the lattice, both electronic and structural properties could still be altered. To determine exactly how much and where Sn sit in the lattice would require for a sophisticated crystallographic investigation. This is clearly beyond the scope of the current investigation.

In summary, comprehensive magnetic neutron diffraction experiments on a Sn-flux grown $BaFe_2As_2$ single crystal have been carried out. The crystal has been observed to undergo a simultaneous antiferromagnetic and tetragonal-orthorhombic structural phase transitions at $T$ ~ 90 K upon cooling down. Despite of an extremely small orthorhombic **a-b** splitting, our results do support a unique magnetic propagation wavevector **k** = (1, 0, 1) and the ordered moment of iron aligned along the longer orthorhombic **a** axis.

We would like to thank Dr. A. Erb for great helps in the sample characterizations.

**Figure captions:**

Figure 1 Temperature dependence of lattice parameters obtained from the Rietveld refinement on high-resolution X-ray powder diffraction. (**a**) Lattice parameter a and b. (**b**) Lattice parameter c and the orthorhombic distortion parameter $P$ (=(a-b)/(a+b)). (**c**) unit cell volume.

Figure 2 (**a**) Triple splitting of the rocking curve of (2, 2, 0) at 3 K; comparing to a sharp rocking curve of (2, 2, 0) at 135 K, measured at HEIDI. (**b**) The temperature dependence of the peak broadening of rocking curves of (4, 0, 0) and (2, 2, 0), but the broadening is not observed at (0, 0, 8), indicating the occurrence of tetragonal to orthorhombic phase transition at 90 K, measured at HEIDI. (**c**) $h$-scan at (-4, 0, 0) at T = 3 K, measured at PANDA, where a slightly asymmetric peak with a shoulder at $h \sim -3.98$ can be seen. The dashed lines are the fittings of Gaussian, the solid lines are the sum of two Gaussian fittings

Figure 3 Magnetic reflections and the temperature dependence measurements. (**a**) A schematic drawing of the ($h$, 0, $l$) plane in the reciprocal space, where the positions of the nuclear and magnetic Bragg reflections are marked by different symbols. (**b**). Rocking curves of $(-1, 0, -3)_M$ measured at 3 K, indicating the occurrence of the antiferromagnetic ordering. The solid lines are the Gaussian fittings. (**c**) The temperature dependence of the intensity of $(-1, 0, -3)_M$, upon cooling (black) and heating (red). All data are taken at PANDA.

Figure 4 (**a**) Comparison of the scans along the $h$-direction at (-1, 0, -3) measured with and without the PG filter. (**b**) Comparison of the intensities of magnetic reflections $(1, 0, 7)_M$ and $(3, 0, 1)_M$ at 3K. $\alpha$ is the angle between the scattering vector and the direction of magnetic moment. The solid line is the Gaussian fitting. (**c**) The magnetic structure of $BaFe_2As_2$ below $T_N$, the directions of iron magnetic moments are indicated by arrows.

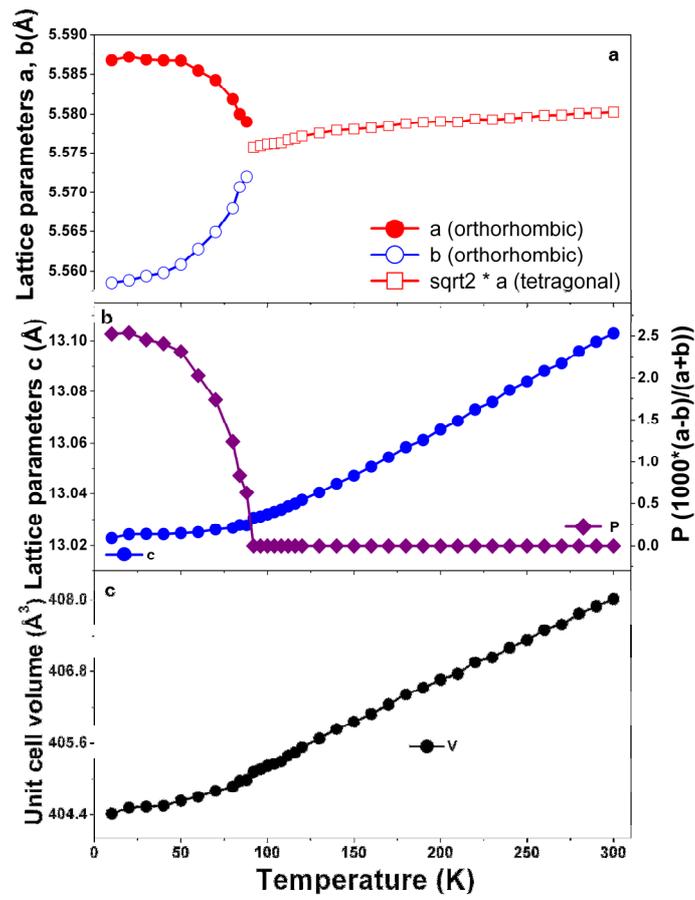

Fig. 1

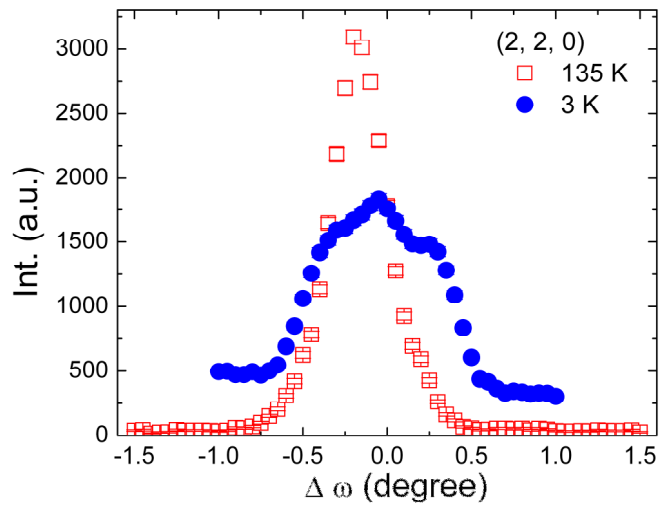

Fig. 2(a)

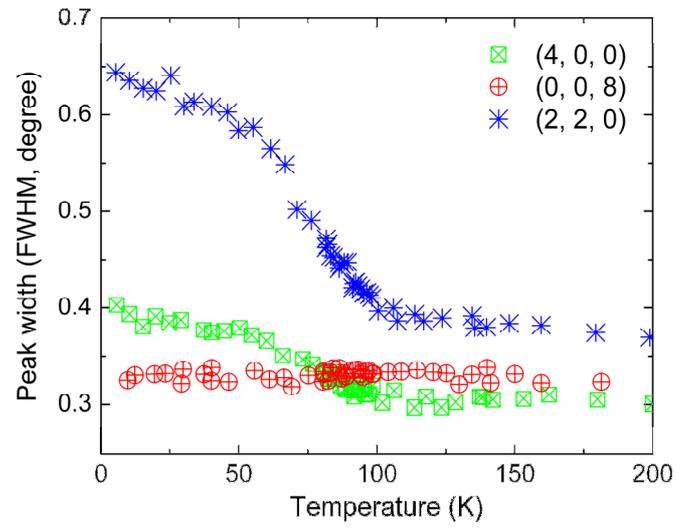

Fig. 2(b)

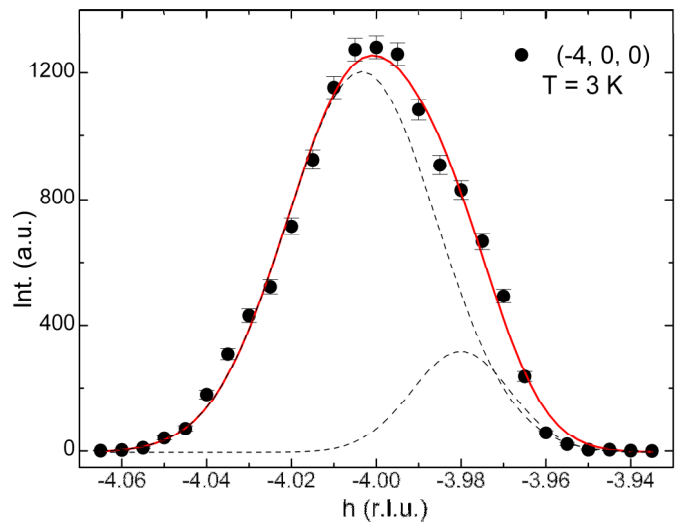

Fig. 2(c)

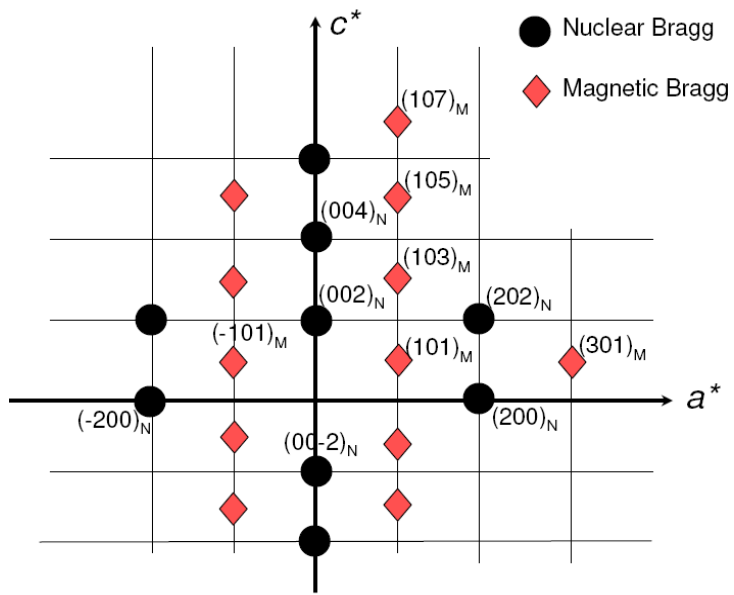

Fig. 3(a)

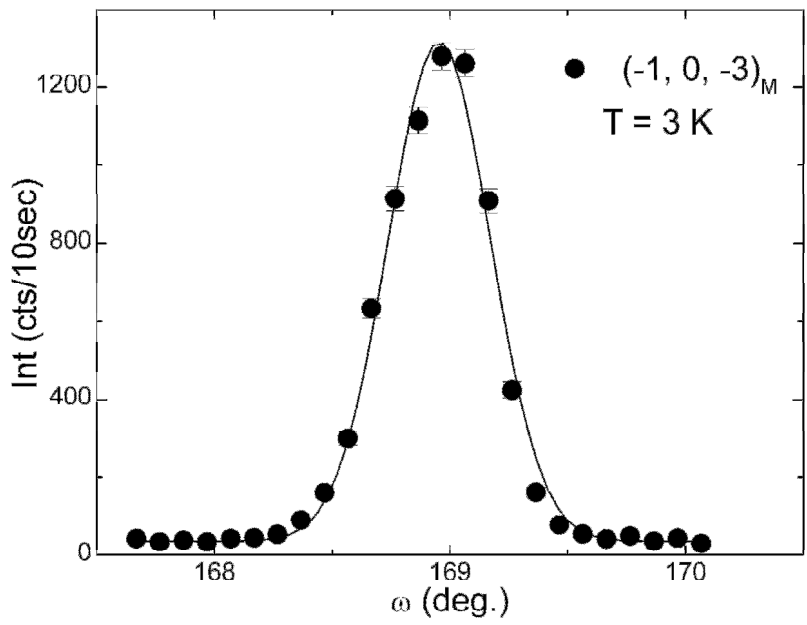

Fig. 3(b)

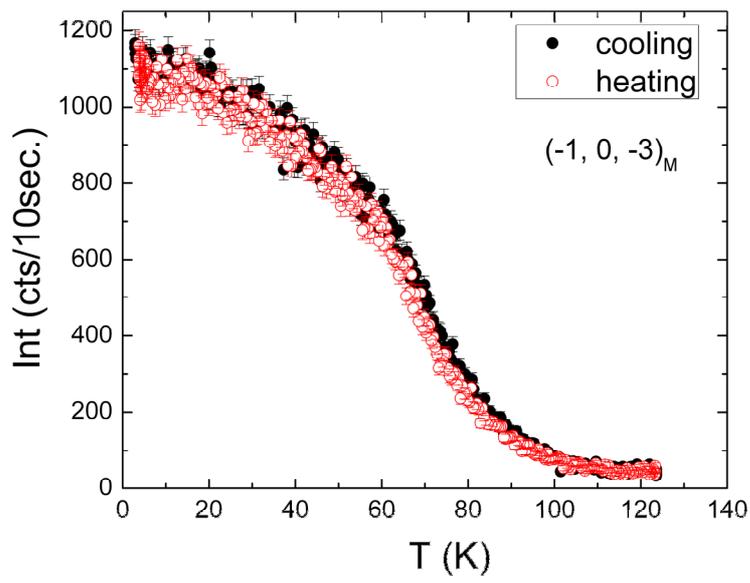

Fig. 3(c)

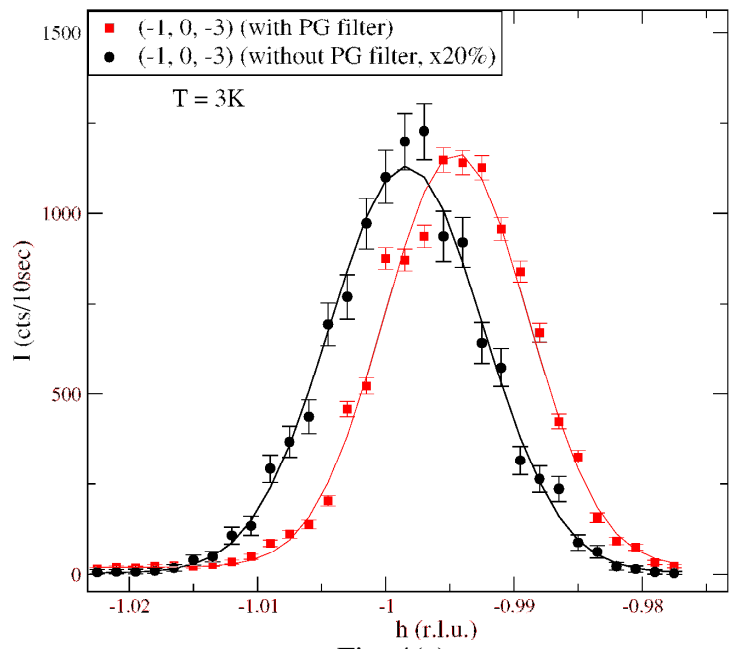

Fig. 4(a)

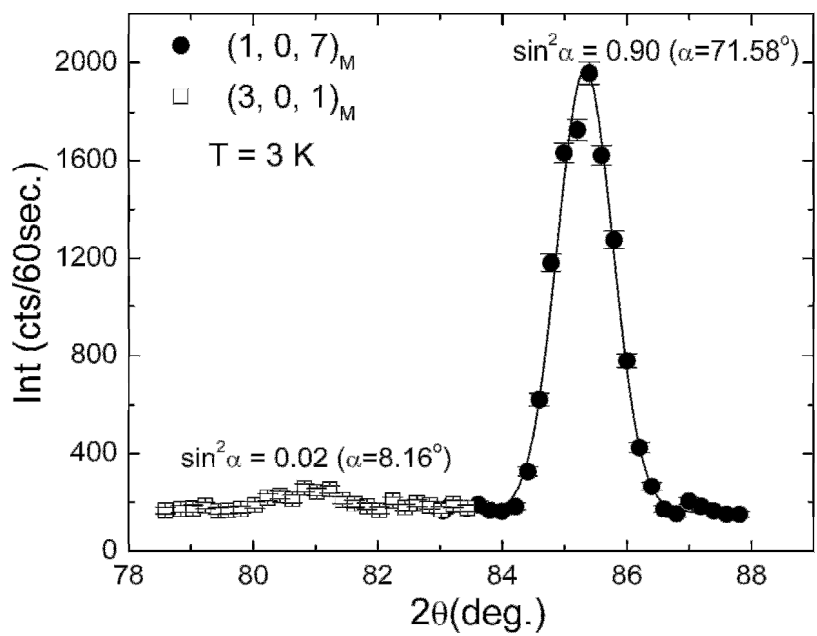

Fig. 4(b)

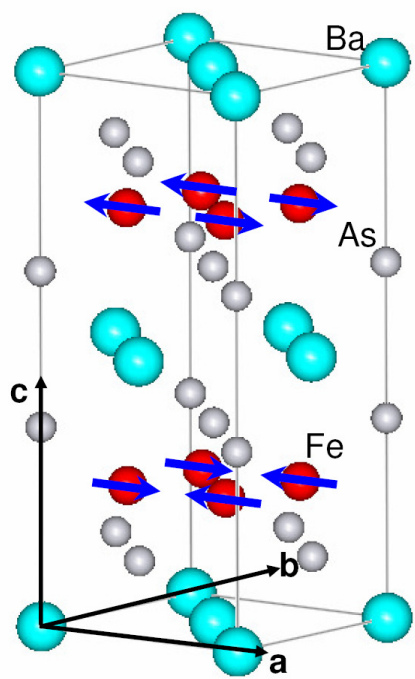

Fig. 4(c)